\documentclass[openacc]{rstransa}%
\usepackage{mathrsfs}
\usepackage{bm}
\usepackage{soul}
\usepackage{mathtools}

\newcommand{\Tr}{\mathop{\mathrm{Tr}} \nolimits}

\newcommand{\op}[1]{\hat{#1}}
\newcommand{\openone}{\leavevmode\hbox{\small1\normalsize\kern-.33em1}}
\newcommand{\W}{\scriptscriptstyle{\mathrm{W}}}

\titlehead{Research}

\begin{document}

\title{Intensity correlations in the Wigner representation}

\author{M. S. Najafabadi$^{1}$, L. L. S\'anchez-Soto$^{1, 2}$, K. Huang$^{3}$, J. Laurat$^{4}$, H. Le Jeannic$^{3}$, and G.~Leuchs$^{1}$}

\address{
$^{1}$Max-Planck-Institut f\"{u}r die Physik des Lichts, 91058Erlangen, Germany\\
$^{2}$Departamento de  \'Optica, Facultad de F\'{\i}sica, Universidad Complutense, 28040~Madrid, Spain\\
$^{3}$State Key Laboratory of Precision Spectroscopy, East China Normal University, 200062 Shanghai, China\\
$^{4}$Laboratoire Kastler Brossel, Sorbonne Universit\'e, ENS-PSL Research University, Coll\`ege de France, 75252~Paris, France}

\subject{Quantum Optics, Quantun Information}


\corres{Gerd Leuchs\\
\email{Gerd.Leuchs@mpl.mpg.de}}

\begin{abstract}
We derive a compact expression for the second-order correlation function $g^{(2)} (0)$ of a quantum state in terms of its Wigner function, thereby establishing a direct link between $g^{(2)} (0)$ and the state's shape in phase space. We conduct an experiment that simultaneously measures $g^{(2)} (0)$ through direct photocounting and reconstructs the Wigner function via homodyne tomography. The results confirm our theoretical predictions. 
\end{abstract}


\begin{fmtext}
 
\section{Introduction}

Phase-space methods were proposed in the early days of quantum mechanics to circumvent some of the conundrums inherent in the conventional Hilbert-space formulation. While Weyl~\cite{Weyl:1928aa} and Wigner~\cite{Wigner:1932aa} laid the groundwork for this approach, it was the contributions of Groenewold~\cite{Groenewold:1946aa} and Moyal~\cite{Moyal:1949aa} that paved the way to formally represent quantum mechanics as a statistical theory in phase space~\cite{Lee:1995aa,Schroek:1996aa,Ozorio:1998aa,Schleich:2001aa,QMPS:2005aa,Baszak:2012aa,Rundle:2021aa}.  This facilitates the emergence of the corresponding classical limit in a more natural and intuitive manner.

The essence of this approach lies in a mapping that associates every operator to a function (known as its symbol)  defined on a smooth manifold with a very precise mathematical structure~\cite{Kirillov:2004aa}.  Unfortunately, this mapping is not unique: a whole family of functions can be consistently assigned to each operator. In particular, 
\end{fmtext}
\maketitle
\noindent 
the quasiprobability distributions, so popular in quantum optics, are just the corresponding symbols of the density operator~\cite{Tatarskii:1983aa,OConnell:1983aa,Balazs:1984aa,Hillery:1984aa,Case:2008aa,Weinbub:2018aa}. For continuous variables, such as position and momentum of a harmonic oscillator (or a single-mode field), the quintessential example that fuelled the interest for this topic, the most common choices are the $P$ (Glauber-Sudarshan)~\cite{Glauber:1963aa,Sudarshan:1963aa}, $W$~(Wigner)~\cite{Wigner:1932aa}, and $Q$~(Husimi)~\cite{Husimi:1940aa,Kano:1965aa} functions, respectively.

These quasidistributions convey complete information about the system, yet each of them corresponds to a different ordering of the  creation and destruction operators. The $P$-representation is utilized to evaluate  normally ordered correlations of field operators, the $Q$ function is associated with antinormal order, and the Wigner function is employed with symmetrically ordered operators.

Direct photodetection corresponds to the expectation values of certain simple products of creation and annihilation operators in normal order~\cite{Glauber:1963aa}. This fundamental aspect underscores the significance of the $P$-representation. The optical equivalence theorem~\cite{Klauder:1966aa} establishes a formal correspondence between expectations of normally ordered operators in quantum optics and expectations of the corresponding $c$-number functions in classical optics. Notably, the $P$-representation stands out from other phase-space densities as it aligns with classical probability when a classical description of the field state exists. However, its behavior for states that are strongly nonclassical can be somewhat problematic, a trade-off necessary for maintaining correspondence with classical optics~\cite{Mandel:1995aa}

The second-order correlation function $g^{(2)}(0)$ is nowadays an essential tool to certify the quantumness of a given state and to distinguish antibunched light sources from classical thermal ones~\cite{Loudon:2000aa}. Photon counting pertains to the realm of $g^{(2)}(0)$ and, consequently, to the $P$ function~\cite{Laiho:2022aa}. 

For numerous reasons~\cite{Ferry:2018aa}, the Wigner function has gained more significance than any other quasiprobability. It can be reconstructed via optical homodyne tomography~\cite{Smithey:1993aa,Breitenbach:1997aa} or directly sampled point-by-point with photon counting and displacement~\cite{Banaszek:1996aa}. Yet it is not directly applicable for evaluating correlation functions, as it is associated with symmetric ordering. The main goal of this paper is to demonstrate how to address this issue and derive quantum correlation functions from moments of the Wigner function.

The plan of this paper is as follows. Section~\ref{sec:pscv} provides a brief overview of the fundamental elements necessary to establish a proper phase-space description of a single-mode quantum field~\cite{rigas2011orbital,seyfarth2020wigner}. In Sec.~\ref{sec:CorrWig}, we demonstrate how to express the correlation function $g^{(2)} (0)$ in terms of the Wigner function, yielding an explicit and simple formula. Section~\ref{sec:exp} details an experiment in which we directly determine $g^{(2)} (0)$ both through direct photon counting and via homodyne detection. Finally, our conclusions are summarized in Sec.~\ref{sec:concl}.

\section{Phase space for quantum continuous variables}
\label{sec:pscv}

In this section we briefly recall the basic ingredients needed to set up a
phase-space description of a single-mode field. The relevant observables are the Hermitian coordinate and momentum quadratures $\op{x}$ and $\op{p}$, with canonical commutation relation $[\op{x}, \op{p}] = i \, \op{\openone}$ (with $\hbar = 1$ throughout).  To avoid technical problems with the unboundedness of $\op{x}$ and $\op{p}$, it is convenient to work with their unitary counterparts
\begin{equation}
  \op{U} (x) = \exp (-i x \, \op{p}) \, ,
  \qquad 
  \op{V} (p) = \exp (- i p \, \op{x}) \, ,
  \label{UV}
\end{equation}
which act on the bases of eigenvectors of position and momentum as 
\begin{equation}
  \label{eq:actbas}
  \op{U} (x^{\prime} ) | x \rangle = | x + x^{\prime} \rangle  \, ,
  \qquad 
  \op{V} (p^{\prime} ) | p \rangle = | p + p^{\prime} \rangle \, ,
\end{equation}
so they represent displacements along the corresponding coordinate axes. The commutation relation is then expressed in the Weyl form~\cite{Galindo:1991fk,Albeverio:2016aa}
\begin{equation}
  \op{V}(p) \op{U}(x) = e^{- ixp} \, \op{U}(x) \op{V}(p) \, .  
  \label{eq:Weyl}
\end{equation}
The infinitesimal version immediately gives the standard commutation relation, but \eqref{eq:Weyl} is more useful in many instances.

In terms of $\op{U}$ and $\op{V}$ a general displacement operator can be introduced as
\begin{equation}
  \label{eq:HWDisp1}
  \op{D} (x, p) =  e^{ i x p/2} \, \op{U} (p)  \op{V}(x)  = 
  \exp[i(p \op{x} -  x \op{p})] \, ,
\end{equation}
with the parameters $(x,p) \in \mathbb{R}^{2}$ labeling phase-space points.  
 
The Fourier transform of the displacement $\op{D} (x, p)$ 
\begin{equation}
  \label{eq:HWkernelDef}
  \op{w} (x, p) = \frac{1}{(2\pi)^2} \int_{\mathbb{R}^2} 
  \exp[-i(p x^\prime - x p^\prime)] \, \op{D} (x^\prime, p^\prime) \, 
  dx^\prime dp^\prime \, ,
\end{equation}
is an instance of a Stratonovich-Weyl quantizer~\cite{Stratonovich:1956kx}. This quantizer can be  easily modified to deal with general $s$-parametrized quasidistributions~\cite{Cahill:1969aa}. One can check that the operators $\op{w} (x, p)$ are a complete trace-orthonormal set that transforms properly under displacements
\begin{equation}
  \label{eq:HWKernelDisp}
  \op{w} (x, p) = \op{D} (x,p) \,\op{w} (0, 0) \,
  \op{D}^\dagger (x, p) \, ,
\end{equation}
where 
\begin{equation}
  \label{eq:HWPar1}
  \op{w}(0,0)= \int_{\mathbb{R}^{2}} \op{D}(x, p) \, dx dp = 2  \op{P}
  \, ,
\end{equation}
and $\op{P} = \int_{\mathbb{R}} | x \rangle \langle - x | \, dx =    \int_{\mathbb{R}} | p \rangle \langle - p | \, d p$ is the parity operator~\cite{Royer:1977aa}.

Let $\op{A}$ be an arbitrary (Hilbert-Schmidt) operator acting on $\mathcal{H}$.  Using the Stratonovich-Weyl quantizer~\eqref{eq:HWkernelDef} we can associate to $\op{A}$ a tempered distribution $a(x, p)$ representing the action of the corresponding dynamical variable in phase space. This is known as the Wigner-Weyl map and reads 
\begin{equation}
  \label{eq:Asy}
  a(x, p)  =   \Tr [ \op{A} \,\op{w}(x,p) ] \, . 
\end{equation}
The function $a (x, p)$ is called the symbol of the operator $\op{A}$. Conversely, we can reconstruct the operator from its symbol through
\begin{equation}
  \label{eq:WigWeyl}
  \op{A} = \frac{1}{(2\pi)^{2}} \int_{\mathbb{R}^{2}} a (x, p) \, 
  \op{w} (x, p ) \, dx dp \, .
\end{equation}

In this context, the Wigner function is nothing but the symbol of the
density matrix $\op{\varrho}$. Therefore, we write
\begin{eqnarray}
  \label{eq:Wigcan}
  &  W_{\varrho}(x, p)  =   \Tr [ \op{\varrho} \, \op{w}(x,p) ] \, , & \nonumber \\ 
  & &  \label{eq:HWWignerDef} \\
  & \op{\varrho}   =   \displaystyle 
  \frac{1}{(2\pi)^{2}} \int_{\mathbb{R}^{2}} \op{w}(x,p) W_{\varrho}(x,p) \, dx dp \, . &
  \nonumber
\end{eqnarray}
For a pure state $| \psi \rangle$, it can  be represented as
\begin{equation}
  \label{eq:1}
  W_{\psi} (x, p) = \frac{1}{2\pi}  \int_{\mathbb{R}}  \exp (i p x^{\prime} ) \,   \psi (x - x^{\prime}/2) \, \psi^{\ast} (x + x^{\prime}/2 )  \, dx^{\prime} \, ,
\end{equation}
which is, perhaps, the most traditional form of writing it.
 
The Wigner function defined in \eqref{eq:HWWignerDef} fulfills all the basic properties required for any good probabilistic description. First,  on integrating $W (x, p)$ over the lines $x_{\theta} = x \cos \theta + p \sin \theta$, the probability distributions of the rotated quadratures $x_{\theta}$ are reproduced
\begin{equation}
  \label{eq:rotquadmad}
  \int_{\mathbb{R}^{2}} W_{\varrho} (x, p) \, \delta (x - x_{\theta} ) \,dx dp =  \langle x_{\theta} | \op{\varrho} | x_{\theta} \rangle \, .
\end{equation}
In particular, the probability distributions for the canonical variables can be obtained as the marginals
\begin{equation}
  \label{eq:HWProps2}
  \int_\mathbb{R}  W_{\varrho} (x, p) \, dp =   
  \langle x | \op{\varrho} | x \rangle   \, ,
  \qquad
  \int_\mathbb{R} W_{\varrho} (x, p) \, dx = 
  \langle p | \op{\varrho} | p \rangle  \, .
\end{equation}
Second, $W_{\varrho} (x, p)$ is translationally covariant, which means that for the displaced state $\op{\varrho}^\prime = \op{D}(x^{\prime}, p^{\prime})$ $\op{\varrho} \, \op{D}^\dagger (x^{\prime}, p^{\prime})$, one has
\begin{equation}
  \label{eq:HWProps3}
  W_{\varrho^\prime} (x, p) = W_{\varrho} (x-x^{\prime}, p-p^{\prime}) \, ,
\end{equation}
so that it follows displacements rigidly without changing its form, reflecting the fact that physics should not depend on the choice of the origin. The same holds true for any linear canonical transformation.

Finally, the overlap of two density operators is proportional to the integral of the associated Wigner functions:
\begin{equation}
  \label{eq:HWProps4}
  \Tr ( \op{\varrho} \,\op{\varrho}^{\prime} ) =
  \int_{\mathbb{R}^2} W_{\varrho} (x, p)  W_{\varrho^{\prime}} (x, p)  \, dx dp \, .
\end{equation}
This property (known as traciality) offers practical advantages, since it allows one to predict the statistics of any outcome, once the Wigner function of the measured state is known. In particular, we have
\begin{equation}
\label{eq:HWPropos5}
\langle A \rangle = \Tr ( \op{\varrho} \, \op{A}) = \int_{\mathbb{R}^2} W_{A} (x, p)  W_{\varrho} (x, p)  \, dx dp \, .
\end{equation}

Coherent states are closely linked with the notion of Gaussian states. The displacements constitute a basic ingredient for their definition: indeed, if we choose a fixed normalized reference state $ |
\psi_{0}\rangle $, we have~\cite{Perelomov:1986kl}
\begin{equation}
  | x, p \rangle = \op{D} ( x, p) \, | \psi_{0} \rangle \, ,  
  \label{eq:defCS}
\end{equation}
so they are parametrized by phase-space points. These states have a number of remarkable properties inherited from those of $\op{D} (x, p)$. The standard choice for the fiducial vector $| \psi_{0} \rangle$ is the vacuum $|0 \rangle $; this guarantees that $\Delta x = \Delta p = 1/\sqrt{2}$, with  $\Delta_{\psi} A = [\langle \op{A}^{2} \rangle - \langle \op{A}  \rangle^{2}]^{1/2}$, and they are  minimum uncertainty states 
\begin{equation}
   \Delta x \, \Delta p =  \frac{1}{2} \, .  
  \label{eq:MUS}
\end{equation}

\section{Correlation functions in the Wigner representation }
\label{sec:CorrWig}

The definition of the normalized second-order correlation function for a single-mode field reads~\cite{Loudon:2000aa}
\begin{equation}
{g^{(2)} (\tau) = \frac{\langle \op{a}^{\dagger}  (t) \op{a}^{\dagger} (t + \tau) \op{a} (t + \tau) \op{a} (t) \rangle}{\langle \op{a}^{\dagger} (t) \op{a} (t) \rangle^{2}}}  \, ,
\end{equation}
where we have introduced the standard creation and annihilation operators
\begin{equation}
\op{a} = \frac{1}{\sqrt{2}} ( \op{x} + i \op{p}) \, , \qquad \qquad
\op{a}^{\dagger} = \frac{1}{\sqrt{2}} ( \op{x} - i \op{p}) \, , 
\end{equation}
with commutation relation $[\op{a}, \op{a}^{\dagger}] = \openone$.

Although all the results we derive in the following can be worked out for $g^{(2)} (\tau)$, things become simpler for the case $\tau = 0$, so that
\begin{equation}
{g^{(2)} (0) = \frac{\langle \op{a}^{\dagger}  \op{a}^{\dagger} \op{a}  \op{a} \rangle}{\langle \op{a}^{\dagger} \op{a} (t) \rangle^{2}}  = \frac{\langle \op{n} (\op{n} - 1) \rangle}{\langle \op{n} \rangle^{2}}} \, ,
\end{equation}
{and $\op{n} = \op{a}^{\dagger} \op{a}$ is the number operator. Note that this of particular interest, since $g^{(2)}(0)$ represents the conditional probability how likely is it to detect a second photon at the same time one photon was already detected. Thus, it is a measure of the temporal photon coincidences, required to distinguish between different
light states.} 

In order to express this correlation function in terms of the Wigner function, we need the notion of Weyl (or symmetric) ordering of operators~\cite{Gosson:2017aa}: for arbitrary powers $k$ and $\ell$, we denote by $\{ \op{a}^{\dagger k } \op{a}^{\ell} \}_{\W}$ the average sum of the $(k+\ell)!/k! \ell!$ different ordered operator products. If $\op{n}_{\W} = \{ \op{a}^{\dagger} \op{a}\}_{\W}$  is the symmetrically ordered number operator, a direct calculation shows that~\cite{Fujii:2004aa}  
\begin{equation}
\begin{aligned}
   \label{Eq:n}
     \op{n}_{\W} & = \op{n} + \tfrac{1}{2}  = \tfrac{1}{2}(\op{x}^{2} + \op{p}^{2}) \, ,\\
     \op{n}_{\W}^{2} & = \op{n}^{2} + \op{n} + \tfrac{1}{2} = \tfrac{1}{4}(\op{x}^{2} + \op{p}^{2})^{2} \,. 
\end{aligned}
\end{equation}
The process can be easily continued to higher powers of the photon number operator and thus our treatment can be extended to any correlation function.

From \eqref{Eq:n}, we immediately get $\op{n}^2 = \op{n}^2_{\W} - \op{n}_{\W}$. Consequently, the correlation function $g^{(2)}(0)$ can be expressed in terms of the symmetrically ordered photon-number operator as 
\begin{equation}
\label{Eq:g^2_sym}
    g^{(2)}(0) =  \frac{\langle \op{n}^2_{\W} \rangle - 2 \langle \op{n}_{\W} \rangle + \tfrac{1}{2}}{\left ( \langle \op{n}_{\W} \rangle -\tfrac{1}{2}  \right )^2}.
\end{equation}
Now, note that the Weyl moments of the number operator can be immediately calculated from the Wigner function:
\begin{equation}
\label{eq:var}
\begin{aligned}
    \langle \op{n}_{\W} \rangle & = \int  \tfrac{1}{2} (x^2 + p^2) \, W_{\varrho}(x, p) \, dx dp \, , \\
    \langle \op{n}^2_{\W} \rangle & = \int \tfrac{1}{4} (x^2 + p^2)^2 \, W_{\varrho} (x, p) \, dx dp \, . 
\end{aligned}
\end{equation}
In this way, we can determine $g^{(2)}(0)$ directly from the Wigner function.  

In what follows, we assume, for simplicity, the case of Gaussian states, whose Wigner function can be compactly written as~\cite{Weedbrook:2012aa} 
\begin{equation}
\label{eq:WigGauss}
W_{\varrho} (\xi)  = \frac{1}{2 \pi \det \bm{\mathsf{V}}} \exp \left [ \tfrac{1}{2} (\xi - \bar{\xi})^{\top} \bm{\mathsf{V}}^{-1}(\xi - \bar{\xi}) \right ] \, .
\end{equation}
Here, we have used the column vector $\xi = (x, p )^{\top}$ (the subscript $\top$ being the transpose), $\bar{\xi} = \langle \xi \rangle$, and the $2 \times 2$ covariance matrix $\bm{\mathsf{V}}$ has elements
\begin{equation}
\mathsf{V}_{ij} = \tfrac{1}{2} \langle \Delta \op{\xi}_{i} \,  \Delta \op{\xi}_{j} +  \Delta \op{\xi}_{j} \,  \Delta \op{\xi}_{i} \rangle
\end{equation}
with $\op{\xi} = (\hat{x}, \hat{p})^{\top}$ and $\Delta \op{\xi}_{i} = \op{\xi}_{i} - \bar{\xi}_{i}$. 

In many interesting instances, the covariance matrix is diagonal so that
\begin{equation}
\bar{\xi} = \begin{pmatrix} 
x_{0} \\
p_{0} 
\end{pmatrix}  \, ,
\qquad \qquad
\bm{\mathsf{V}} = \begin{pmatrix} 
(\Delta \hat{x})^{2} & 0 \\
0 & (\Delta \hat{p})^{2}
\end{pmatrix} \, .
\end{equation}
The corresponding Wigner function then reduces to
\begin{equation}
W_{\varrho} (x,p) = \frac{1}{2 \pi \Delta x \, \Delta p}
\exp \left [  - \frac{(x-x_{0})^{2}}{2 (\Delta x)^{2}}- \frac{(p-p_{0})^{2}}{2 (\Delta p)^{2}}\right ] \,. 
\end{equation}

For a coherent state the Weyl moments \eqref{eq:var} can be directly computed; the result is 
\begin{equation} 
\langle \op{n}_{\W} \rangle = \tfrac{1}{2}|\bar{\xi}|^{2} + \tfrac{1}{2} , \qquad
\langle \op{n}_{\W}^{2} \rangle = |\bar{\xi}|^{4} +  2 |\bar{\xi}|^{2}+ \tfrac{1}{2} , 
\end{equation}
where $|\bar{\xi}|^{2} = x_{0}^{2} + p_{0}^{2}$. Using \eqref{Eq:g^2_sym}, we get (see Appendix)
\begin{equation}
g^{(2)}_{\mathrm{coh}} (0) = 1 \, .
\end{equation}

\begin{figure}
    \centering
    \includegraphics[width=.60\columnwidth]{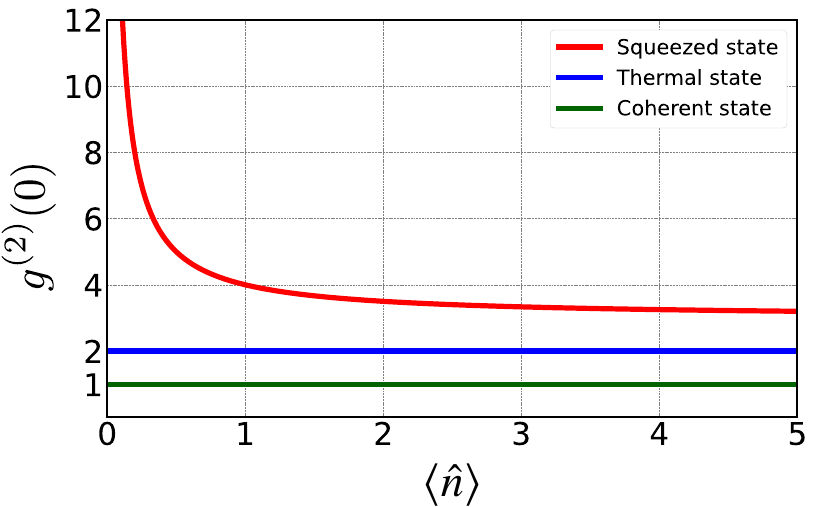}
    \caption{Second-order correlation function as a function of the mean photon number for the three typical states considered in this work.  Note, that the red curve is for pure squeezed states and not for attenuated squeezed states.}
    \label{fig:Figure_g2_N}
\end{figure}

For a thermal state (see Appendix)
\begin{equation}
\op{\varrho} = \sum_{n=0}^{\infty} \frac{\bar{n}^{n}}{(1+\bar{n})^{n}} |n \rangle \langle n| \, ,
\end{equation}
where $\bar{n} = [\exp(\hbar \omega/k_{\mathrm{B}} T) - 1]^{-1}$ is the average number of photons, we have $\Delta x = \Delta p =\sqrt{2 \bar{n} + 1}$  and we can check that $\langle \op{n}^{2}_{\W} \rangle = 2\langle \op{n}_{\W} \rangle^2$, so that 
\begin{equation}
\label{eq:g2th}
g^{(2)}_{\mathrm{th}}(0) = 2 \, . 
\end{equation}

Both, thermal and coherent light are characterized by a two-dimensional symmetric noise. The distinctly different $g^{(2)}(0)$ for the coherent state is a result of its uncertainty being at the quantum limit given by $\langle \op{n}_{\W} \rangle = 1/2$  for $\langle \op{n} \rangle = 0$ and is related to the non-commuting field operators.

As a final example of Gaussian states, we consider squeezed states, for which  $(\Delta x)^{2} = e^{-r}$ and $(\Delta p)^{2} = e^{r}$, where $r$ is the squeezing parameter~\cite{Leuchs:1986aa,Loudon:1987aa}. Now, we have 
\begin{equation}
\langle \op{n}_{\W} \rangle  = \tfrac{1}{2} ( e^{r} + e^{-r} ) \, , \qquad \langle \op{n}^{2}_{\W} \rangle  =  3 \langle \op{n}_{\W} \rangle^{2} - \tfrac{1}{4} \, .
\end{equation}
From here, we obtain that
\begin{align}
\label{eq:g2sq}
 g^{(2)}_{\mathrm{sq}}(0)= 3 + \frac{1}{\langle \op{n} \rangle} \, .
\end{align}

The different values 1, 2 and 3 of $g^{2}(0)$  corresponding to a coherent state, a thermal state, and a highly squeezed vacuum have been interpreted as originated from the different dimensionality of these states in phase space~\cite{Leuchs:2015aa,Leuchs:2015ab}. This establishes a direct link between $g^{2}(0)$ and the shape of the state in phase space. 

Figure~\ref{fig:Figure_g2_N} depicts $g^{(2)}(0)$ for coherent, thermal, and squeezed states as a function of $\langle \op{n} \rangle = \langle \op{n}_{\W} \rangle -1/2$. As it is clear, in the case of a squeezed vacuum, as the average photon number approaches to zero, $g^{(2)}(0)$ diverges. Under attenuation, squeezed states become mixed, altering their variance but their $g^{(2)}(0)$ does not change. In contrast, coherent and thermal states remain coherent and thermal, respectively, under attenuation. Consequently, their $g^{(2)}(0)$ is not only constant under attenuation but also independent of the average photon number. 
 
\section{Experiment}
\label{sec:exp}

\begin{figure}[t]
    \centering
    \includegraphics[width=.70\columnwidth]{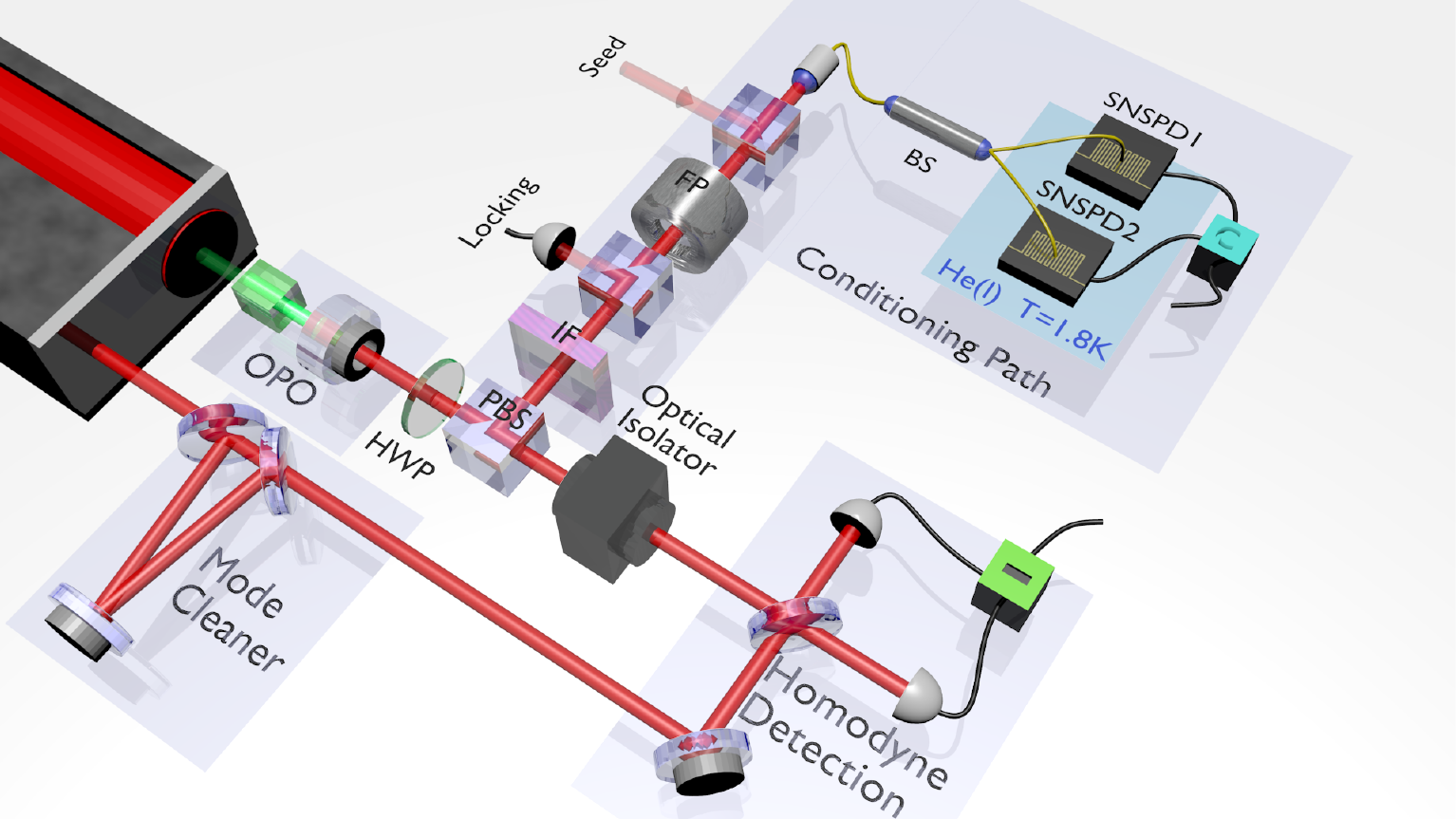}
    \caption{Experimental setup (for more details see Ref.~\cite{Huang:2015aa}): a type-II OPO is continuously pumped with a 532~nm Nd:YAG laser. The output modes are mixed using a polarized beam splitter (PBS) and a half-wave plate (HWP). A two-photon detection is implemented on one output of the PBS via multiplexed superconducting nanowire single-photon detectors (SNSPDs) after frequency filtering. The other output is analyzed via a homodyne detection for full quantum state tomography.}
    \label{fig:experiment}
\end{figure}

We check our theory with an experiment able to perform simultaneously both measurements; direct and homodyne. The setup is sketched in Fig~\ref{fig:experiment} and has been detailed in Ref.\cite{Huang:2015aa}. We use a type-II phase-matched KTP triply-resonant optical parametric oscillator, pumped far below threshold by a continuous-wave Nd:YAG laser at 532~nm. The output polarization modes can be mixed using a half-wave plate (HWP) and a polarizing beam splitter (PBS). One output of the PBS is frequency filtered with an interferential filter and a cavity. This path is then split on a fiber coupler BS, and the outputs are detected via two superconducting nanowire single-photon detectors (SNSPDs)~\cite{Jeannic:2016aa}. This configuration enables us to check the second-order correlation function. The other output of the PBS is measured via homodyne detection, thus enabling to reconstruct the density matrix and the associated Wigner function. 

For each angle of the HWP, we can thereby record $g^{(2)}(\tau)$ and the density matrix of its complementary. However, the two recorded states only differ by a rotation in phase space. Therefore their quadrature variances and $g^{(2)}(\tau)$ are the same. So, we can measure these two quantities simultaneously and separately on the two output ports of the PBS.

When the HWP is at $0^{\circ}$ it corresponds to the perfect separation of the two orthogonally polarized modes. The resulting state is a twin beam, exhibiting EPR entanglement~\cite{Chekhova:2015aa}:
\begin{equation}
|\psi \rangle = \sqrt{1 - |\lambda|^{2}} \sum_n \lambda^{n} |n \rangle_{s} |n \rangle_{i} \, ,
\end{equation}
where $\lambda = e^{i \phi_{0}} \tanh r$, with $\phi_{0}$ being the pump field phase, and the subscript $s$ and $i$ refer to the signal and idler modes, respectively. For this angle, the output is thus made of two correlated thermal states, and we have $g_{\mathrm{th}}^{(2)} (0) = 2$ in each of the two output ports of the PBS. 

When increasing the angle, a squeezed vacuum is produced with $g_{\mathrm{sq}}^{(2)} (0) $ given by \eqref{eq:g2sq}. Through this basis change we introduce a continuous transition in Fig.~\ref{fig:Figure_g2_N} in the vertical direction from thermal to squeezed statistics.

When the polarization basis is rotated by $45^{\circ}$, the signal and idler modes can be rewritten as 
\begin{equation}
\op{a}_{s} = \op{a}_{1}  + i \op{a}_{2} \, , \qquad 
\op{a}_{i} = \op{a}_{2}  - i \op{a}_{1} \, ,
\end{equation}  
so that we have two uncorrelated squeezed vacuum states on each of the spatial modes 1 and 2:
\begin{equation}
|\psi \rangle \propto \left ( \sum_{n} c_{2n} | 2n \rangle_{1} \right )
\left ( \sum_{n} c_{2n} | 2n \rangle_{2} \right )\, . 
\end{equation}  
This occurs when the angle of the HWP is $22.5^{\circ}$. In consequence, depending on the  HWP we can  either generate EPR-entanglement, leading to thermal states in each spatial mode, or to decorrelate the modes, which leads to independent squeezed vacua on each mode. This situation corresponds to the transition from two independent single-mode squeezers to one two-mode squeezer.

\begin{figure}[t]
    \centering
    \includegraphics[width=.65 \columnwidth]{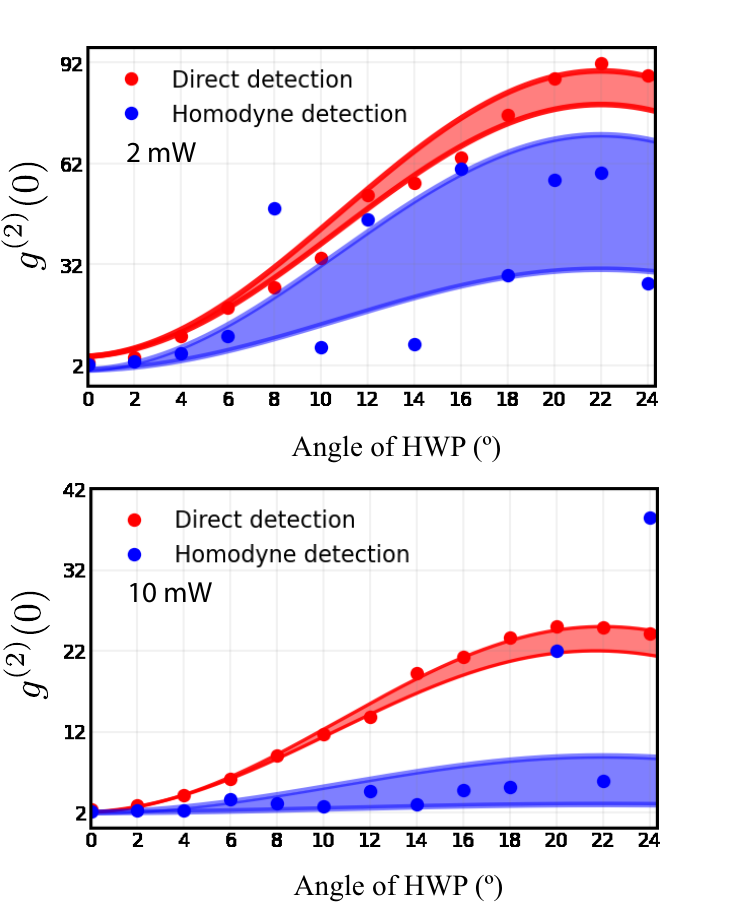}
    \caption{Values of $g^{(2)}(0)$ as a function of the angle of the wave plate, for two different input powers indicated in the insets. In each figure we plot the results obtained from both direct photon counting and via the Wigner function reconstructed from homodyne detection. The  shadows indicate the errors associated with both methods.}  
    \label{fig:compare}
 \end{figure}

The filtered path enables to measure the second order correlation function. To achieve this, we acquire the time of the two single-photon detections within an acceptance window of 50~ns. The distribution of the photon coincidences depending on the delay $\tau$, normalized by the uncorrelated coincidences, gives $g^{(2)}(\tau)$.  Two datasets were acquired for 2~mW and 10~mW of pump power, corresponding respectively to 0.7 and 1.6 dB of squeezing. The experimental values correspond to the blue points in Fig.~\ref{fig:compare}. The solid line represents the theoretical fitting of the experimental data points with the model $f (\theta) = a \sin[(b+\theta)\pi/45)]^2+c$, where $\theta$ is the angle of the HWP, and $\{a, b, c \}$ are the fitting parameters. The shaded area is the confidence region of the model.    

At $\theta=0^{\circ}$, we observe a value $g^{(2)}(0) \simeq 2.2$ for both powers, slightly higher than theoretical value. At $\theta = 22.5^{\circ}$, the values are $g^{(2)}(0) \simeq 27.5$ and $g^{(2)}(0) \simeq 90$ for 2~mW and 10~mW, respectively. This is consistent with mean photon numbers of $0.0115$ and $0.004$, respectively.

With our homodyne detection setup, for the same HWP angle, we can also reconstruct the full density matrix of the state and derive the associated Wigner function using standard methods. We can thus witness the transition from a squeezed state, where one quadrature has smaller fluctuations than the other, to a thermal state. 

\begin{figure}[t]
    \centering
    \includegraphics[width=0.85 \columnwidth]{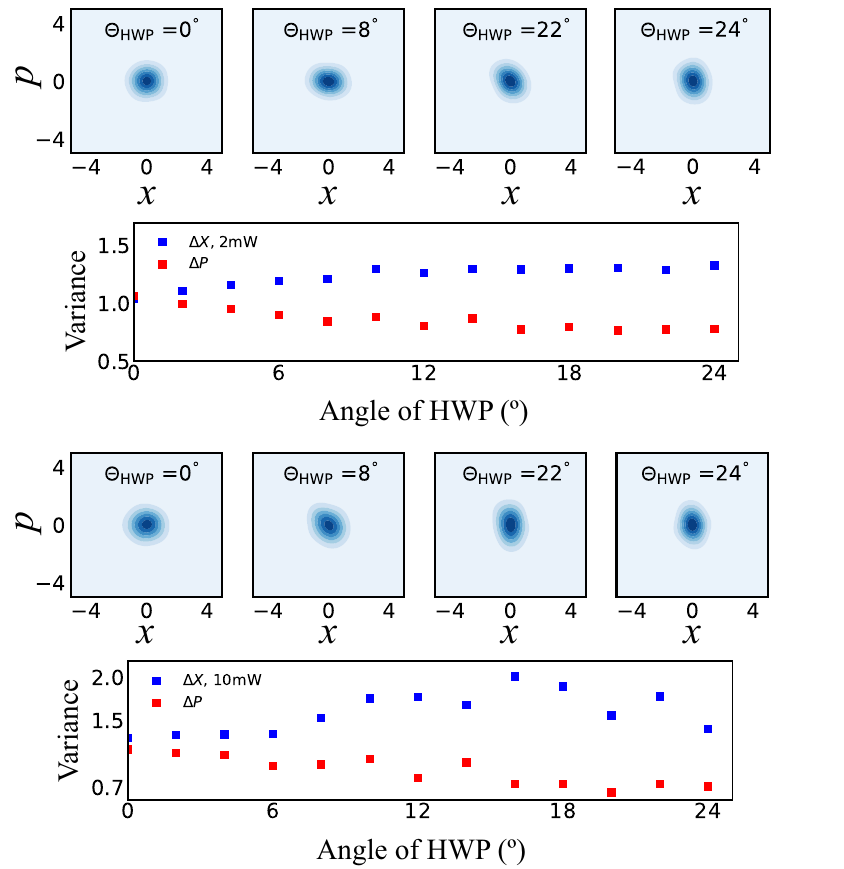}
    \caption{Contour plots of the reconstructed Wigner functions for several HWP angles showing the transition from a thermal state corresponding to the angle $0^{\circ}$ to a squeezed vacuum state corresponding to $22.5^{\circ}$. We also include the variances derived from the reconstruction. The top panel correspond to 2 mW and the bottom panel to 10 mW.}
    \label{fig:Fig3_Wigner_Exp}
\end{figure}

To determine the quadrature variances, we fit the Wigner functions by a Gaussian distribution model  
\begin{equation}
\label{eq:Wig}
    W(x,p) = W_{0} \exp[ - F(x,y) ] \, ,
\end{equation}
where
\begin{align}
\nonumber
  F(x,y) & = \frac{\left [(x-x_0) \cos\theta +(p-p_0) \sin\theta \right ]^2}{2(\Delta x)^2} 
  + \frac{\left[- (x-x_0) \sin \theta + (p-p_0) \cos \theta \right ]^2}{2(\Delta p)^2} \, .
\end{align}

The reconstructed Wigner functions for some values of the angle $\theta$ of the HWP are given in Fig.~\ref{fig:Fig3_Wigner_Exp}, as well as the resulting variances for each quadrature.  Using these variances we can now calculate the corresponding values of $g^{(2)} (0)$; they are plotted in Fig.~\ref{fig:compare}. The blue shadow indicates the confidence regions. 

While the directly measured $g^{(2)}(0)$ nicely follows the theoretically expected curve when rotating the two-mode basis, the values inferred from the Wigner function are sometimes significantly different. To better understand these discrepancies, it is important to note that in the expression \eqref{eq:g2sq} for pure squeezed states, both the denominator and the numerator approach zero as the mean photon number approaches zero. This can result in very large ratios without a definite limit.  In the same limit, the corresponding Wigner function differs only slightly from that of the vacuum. Thus, even minor noise or imperfections in the experimental setup can have a drastic effect on the inferred data.

It is worth noting that starting with a pure squeezed state and then decreasing the mean photon number by attenuation leads to values of $g^{(2)}(0)$ no longer following this curve for pure squeezed states. Instead, the new $g^{(2)}(0)$ remains constant on a horizontal line. Consequently, two Wigner functions (pure and mixed), both very close to the vacuum state, correspond to significantly different values of $g^{(2)}(0)$.

Appreciating the numerical and experimental challenges, the results are nonetheless encouraging. We plan to redo the experiment with an emphasis on much larger mean photon numbers, where the corresponding Wigner functions are significantly different from each other. Additionally, we will repeat the experiments reported here in the regime of very low mean photon numbers with the utmost care to identify all parameters affecting $g^{(2)}(0)$ in this delicate regime.

An interesting application is the experimental determination of the overall losses by comparing the directly measured $g^{(2)}(0)$ with the homodyne measurement of the squeezed quadrature. The measured  variance $(\Delta x)^{2}_{\mathrm{meas}}$ is affected by losses. The $g^{(2)}(0)$ is not affected by losses. Therefore, we can deduce the symmetrically ordered mean photon number for the underlying unattenuated squeezed state: 
\begin{equation}
\langle \hat{n}_{\W} \rangle_{\mathrm{sq}} = \frac{1}{g_{\mathrm{sq}}^{(2)}(0)- 3}  + \frac{1}{2} 
\end{equation}
from the $g_{\mathrm{sq}}^{(2)} (0)$ measurement. The quadrature variance, $(\Delta x)^{2}_{\mathrm{sq}}$, of the underlying squeezed states without losses can be deduced from $\langle \hat{n}_{\W} \rangle_{\mathrm{sq}} = [(\Delta x)^{2}_{\mathrm{sq}} + 1 / (\Delta x)^{2}_{\mathrm{sq}}]/2$. Using the  relation between the quadrature variances with and without losses, $(\Delta x)^{2}_{\mathrm{meas}}$ and $(\Delta x)^{2}_{\mathrm{sq}}$, one can determine the overall losses. This procedure with directly detecting photodiodes in the homodyning channel is somewhat in the spirit of the proposal by Klyshko~\cite{Klyshko:1977aa} for click detectors. For more recent work see~\cite{Brida:2006aa,Agafonov:2011aa}.

To conclude, we note that we have only considered Gaussian states. The method works though for non-Gaussian states, for which the Wigner function can take on non-positive values. This indicates that different quadratures cannot be jointly measured, reflecting the inherent quantumness of those states. In such cases, the vacuum state often has a significant contribution to the overall quantum state, especially for weak light fields. The experimental limitations due to this strong vacuum contribution have already been discussed~\cite{Hong:2017aa}.

\section{Concluding remarks}
\label{sec:concl}

Photon counting and the ensuing photon correlation properties is the main experimental technique in the discrete-variable approach to quantum optics. On the other hand, the continuous-variable approach is based in quasiprobabilities, the Wigner function being the most conspicuous of them. These two worlds, providing both complete information, do not talk much to each other. In this paper, we have shown how one can pass from one to the other in a crystal-clear manner. Using a unique experimental setup that can work in both worlds, we have shown this equivalence.

\ack{It is our pleasure to dedicate this article to Rodney Loudon, who will be remembered as a pioneer of quantum optics. One of us (GL) fondly and thankfully remembers the period about 25 years ago, when Rodney Loudon visited Erlangen frequently as an Awardee of Alexander von Humboldt Foundation. }

\appendix
\section{Intensity correlation function for Gaussian states}
Here we derive an explicit expression for $g^{(2)}(0)$ using the Weyl moments \eqref{eq:var} for Gaussian states. We take $x= x_0 + \Delta x$ and $p=p_0+ \Delta p$, so that we get  
\begin{equation}
\label{Eq:nW}
\begin{aligned}
    \langle \op{n}_{\W} \rangle & = \tfrac{1}{2} [x_{0}^2 + p_{0}^2+ (\Delta x)^2 + (\Delta p)^2 ], \\
    \langle \op{n}_{\W}^2 \rangle &  = \tfrac{1}{4} \{ x_0^4 + p_0^4 + 3 (\Delta x)^4 + 3 (\Delta p)^4 + 2 (\Delta x)^2 (\Delta p)^2 \\
    & + 2 x_0^2 [ (\Delta p)^2  + 3 (\Delta x)^2 ] +  2 p_0^2 [ (\Delta x)^2  + 3 (\Delta p)^2 ]  \} \,. 
    \end{aligned}
\end{equation}
The same results can be derived using the explicit form of the Wigner function \eqref{eq:WigGauss} and performing the resulting integrals.  
Replacing these values in the general expression  \eqref{Eq:g^2_sym}, we can obtain the values for the states considered here. 
 For a coherent state $\Delta x = \Delta p = 1/\sqrt{2}$ and then
\begin{equation} 
\langle \op{n}_{\W} \rangle = \tfrac{1}{2}|\bar{\xi}|^{2} + \tfrac{1}{2} , \qquad
\langle \op{n}_{\W}^{2} \rangle = |\bar{\xi}|^{4} +  2 |\bar{\xi}|^{2}+ \tfrac{1}{2} , 
\end{equation}
with $\bar{\xi} = x_0^2 + p_0^2$. Hence, we get
\begin{equation}
    \begin{aligned}
        g_{\mathrm{coh}}^{(2)} (0)= 1.
    \end{aligned}
\end{equation}

For a thermal state, we can take $x_0= p_0 = 0$ and $\Delta x = \Delta p = \sqrt{2 \bar{n} + 1} $, which leads:
\begin{equation} 
\langle \op{n}_{\W} \rangle = 2 \bar{n} + 1 , \qquad
\langle \op{n}_{\W}^{2} \rangle = 2 ( 2 \bar{n} + 1)^{2}  \, ,
\end{equation}
so that
\begin{equation}
    \begin{aligned}
        g_{\mathrm{th}}^{(2)} (0)= 2.
    \end{aligned}
\end{equation}

Finally, for the squeezed state we can take without loss of generality $x_0 = p_0 =  0$, and  $(\Delta x)^2 =\tfrac{1}{2}e^{-r}$ and $(\Delta p)^2 =\frac{1}{2}e^{r}$. Now, 
\begin{equation}
\langle \op{n}_{\W} \rangle  = \tfrac{1}{4} ( e^{r} + e^{-r} ) \, , \qquad \langle \op{n}^{2}_{\W} \rangle  =  3 \langle \op{n}_{\W} \rangle^{2} - \tfrac{1}{4} \, ,
\end{equation}
and so
\begin{equation}
    \begin{aligned}
        g^{(2)}_{\mathrm{sq}} (0) = 3 + \frac{1}{\langle n_W \rangle -\frac{1}{2}} = 3 + \frac{1}{\langle n \rangle } .
    \end{aligned}
\end{equation}


\end{document}